\documentclass[10pt,letterpaper,twocolumn]{article} %% two column, final layout

\usepackage{ol2}
\usepackage[draft]{hyperref}
\usepackage{amsmath}

\begin{document}

\twocolumn[ %% activate for two-column option

\title{Can one passively phase lock 25 fiber lasers?}

%% For REVTeX it is possible to automate superscript and e-mail callouts with the superscriptaddress option; see REVTeX4 documentation.

\author{Moti Fridman$^*$, Micha Nixon, Nir Davidson and Asher A. Friesem}

\address{
Weizmann Institute of Science, Dept. of Physics of Complex Systems,
Rehovot 76100, Israel \\ $^*$Corresponding author:
moti.fridman@weizmann.ac.il }

\begin{abstract}
Yes, it is possible to phase lock 25 fiber lasers but only for a
short time. Our experiments on passively phase locking
two-dimensional arrays of coupled fiber lasers reveal that the
average phase locking level of 25 lasers is low ($20\%-30\%$) but
can exceed $90\%$ on rare instantaneous events. The average phase
locking level was found to decrease for larger number of lasers in
the array and increase with the connectivity of the array.
\end{abstract}


 ] %% activate for two-column option

\noindent

Phase locking of many fiber lasers has been attracting much
attention, mainly because it enables their coherent addition so as
to obtain high output brightness beyond the limit of a single fiber
laser~\cite{no_more_siegman, no_more_shirakawa, no_more_rothenberg,
no_more_shakir, FanReview, motiCommonLM, shakir, FanFluctutuate,
FridmanOL, michaTime}. In order to achieve phase locking, the fiber
lasers must have at least one longitudinal mode that is common to
all~\cite{motiCommonLM}. Unfortunately, since the length of each
fiber laser cannot be accurately controlled, the probability for
having such common longitudinal modes rapidly decreases as the
number of coupled fiber lasers increases. Indeed, a theoretical
upper limit of 8 to 12 fiber lasers for efficient phase locking was
predicted~\cite{no_more_siegman, no_more_shirakawa,
no_more_rothenberg, no_more_shakir}. Experimentally, none of the
approaches for passive phase locking or coherent combining of
separate fiber lasers have exceeded this theoretical limit (see e.g.
~\cite{FanReview}).

In this letter we present a new configuration for coupling and phase
locking up to 25 fiber lasers, each with a different cavity length.
With this configuration we investigated in some detail the phase
locking level as a function of the number of fiber lasers and the
connectivity between them. We show that the average phase locking
level reduced significantly for more than ten fiber lasers, in
agreement with the theoretical predictions of
~\cite{no_more_siegman, no_more_shirakawa, no_more_rothenberg,
no_more_shakir}. However, the instantaneous phase locking level
fluctuates rapidly and on some rare occasions it can exceed $90\%$
even for 25 fiber lasers. This is because the cavity lengths of the
fiber lasers typically fluctuate~\cite{FanFluctutuate}, mainly due
to thermal and acoustic variations, so on an instantaneous basis all
lengths can match.

\begin{figure*}[h]
\centerline{\includegraphics[width=14cm]{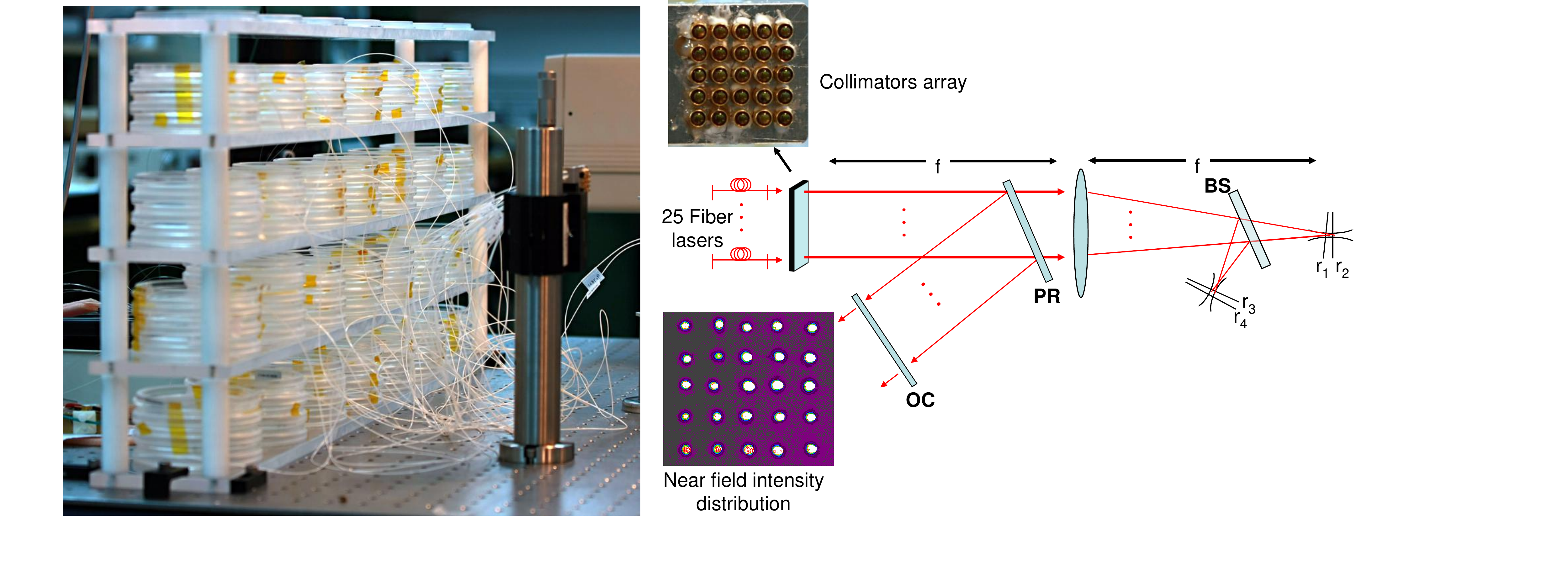}}
\caption{\label{setup}Experimental configuration for phase locking
25 fiber lasers. BS - Beam splitter, OC - output coupler, PR -
partially reflecting mirror.}
\end{figure*}

Our experimental configuration for phase locking up to 25 fiber
lasers is presented schematically in Fig.~\ref{setup}. It is
composed of 25 fiber lasers as shown in the left inset. Each fiber
lasers was comprised of $1.5m \pm 0.2m $ of Ytterbium doped double
clad fiber, a high reflection fiber Bragg grating (FBG) that served
as the rear mirror, and a low reflection FBG ($5\%$) that served as
the front mirror; the spectral bandwidth of either FBGs was $1nm$
with a central wavelength of $1070nm$. Each laser was pumped through
the rear FBG with a $975nm$ stabilized diode laser of $100mW$, and
the front end was attached to a collimator. The collimators of all
the 25 lasers were accurately aligned to form a $5X5$ square array,
as shown in the upper inset, to obtain 25 parallel beams with
parallelism better than $0.1mrad$. The separation between adjacent
lasers was $3.6mm$.

The coupling between the fiber lasers was achieved by means of four
coupling mirrors denoted as $r_1$, $r_2$, $r_3$ and $r_4$ with
reflectivity of $40\%$ for $r_1$ and $r_3$ and reflectivity of
$100\%$ for $r_2$ and $r_4$. All the coupling mirrors were located
close to the focal plane of a focusing lens with $500mm$ focal
length. Since there was only enough space for only a pair of mirrors
within the Rayleigh range of the focussing lens, we inserted a
$50\%$ beam splitter to obtain another focal plane where we placed
another pair of mirrors. Finally, we directed about $10\%$ of the
light with a partially reflecting mirror (PR), from the collimators
array, towards an output coupler (OC), which was placed at a
distance equivalent to that of the focal plane of the lens. The OC
reflected part of the light from each laser back onto itself with
the same delay as the light that is coupled from the other lasers.
Such an arrangement ensures that although the coherence length of
each laser is shorter than the distance to the coupling mirrors,
phase locking can still occur~\cite{michaTime}. The near-field
intensity distribution at the output coupler is presented at the
lower inset of Fig.~\ref{setup}. The intensities of all lasers were
about $50mW$ and remained relatively constant.

\begin{figure}[htb]
\centerline{\includegraphics[width=8cm]{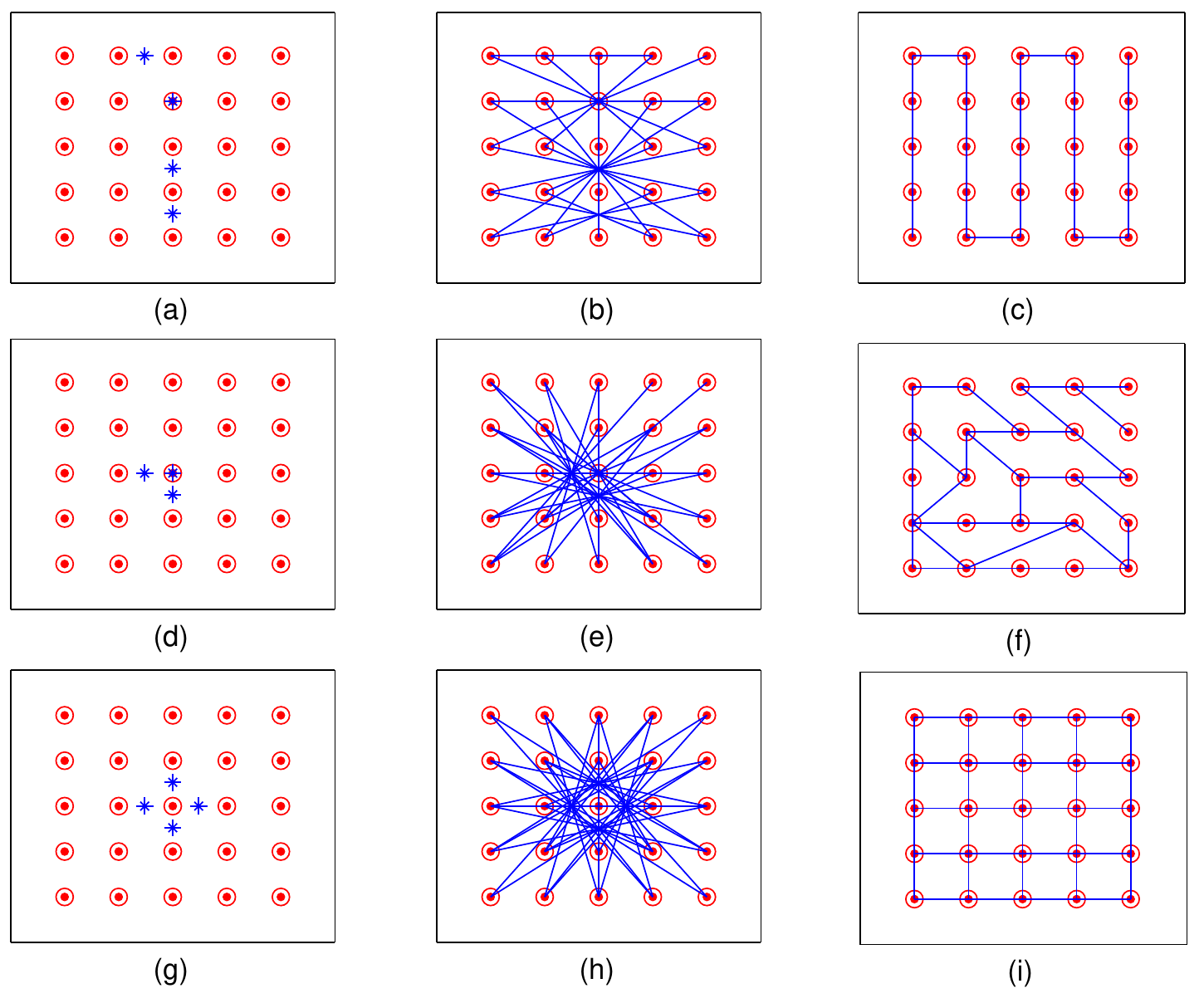}}
\caption{\label{connectAll}Connectivities for three different
orientations of the coupling mirrors.  (a),(d) and (g) show the self
reflection points of each mirror as the blue stars; (b), (e) and (h)
show the corresponding coupling connections; (c), (f) and (i) show
the corresponding coupling connections after rearranging the lasers
locations.}
\end{figure}

Now, by controlling the orientation of the coupling mirrors $r_1$,
$r_2$, $r_3$ and $r_4$, we could realize variety of connectivities
for the fiber lasers array. Each coupling mirror connects pairs of
lasers that are symmetric around the self reflecting point of the
mirror. Representative connectivities for the array of 25 fiber
lasers (denoted as red circles) with three different orientations of
the four coupling mirrors are presented in Fig.~\ref{connectAll}.
Figures~\ref{connectAll}(a),(d) and (g) show the self-reflecting
point of each coupling mirror on the array (denoted as blue stars).
Figures~\ref{connectAll}(b), (e) and (h) show the corresponding
couplings connections that are imposed by the coupling mirrors.
Figures.~\ref{connectAll}(c), (f) and (i) illustrate more clearly
the corresponding coupling connection after we rearranged the lasers
locations in the array. As evident, there is great flexibility for
the coupling connections. These include the effective 1D
connectivity  as shown in Fig~\ref{connectAll}(c), 2D connectivity
as shown in Fig~\ref{connectAll}(i) and an intermediate connectivity
as shown in Fig~\ref{connectAll}(f).

\begin{figure}[htb]
\centerline{\includegraphics[width=8cm]{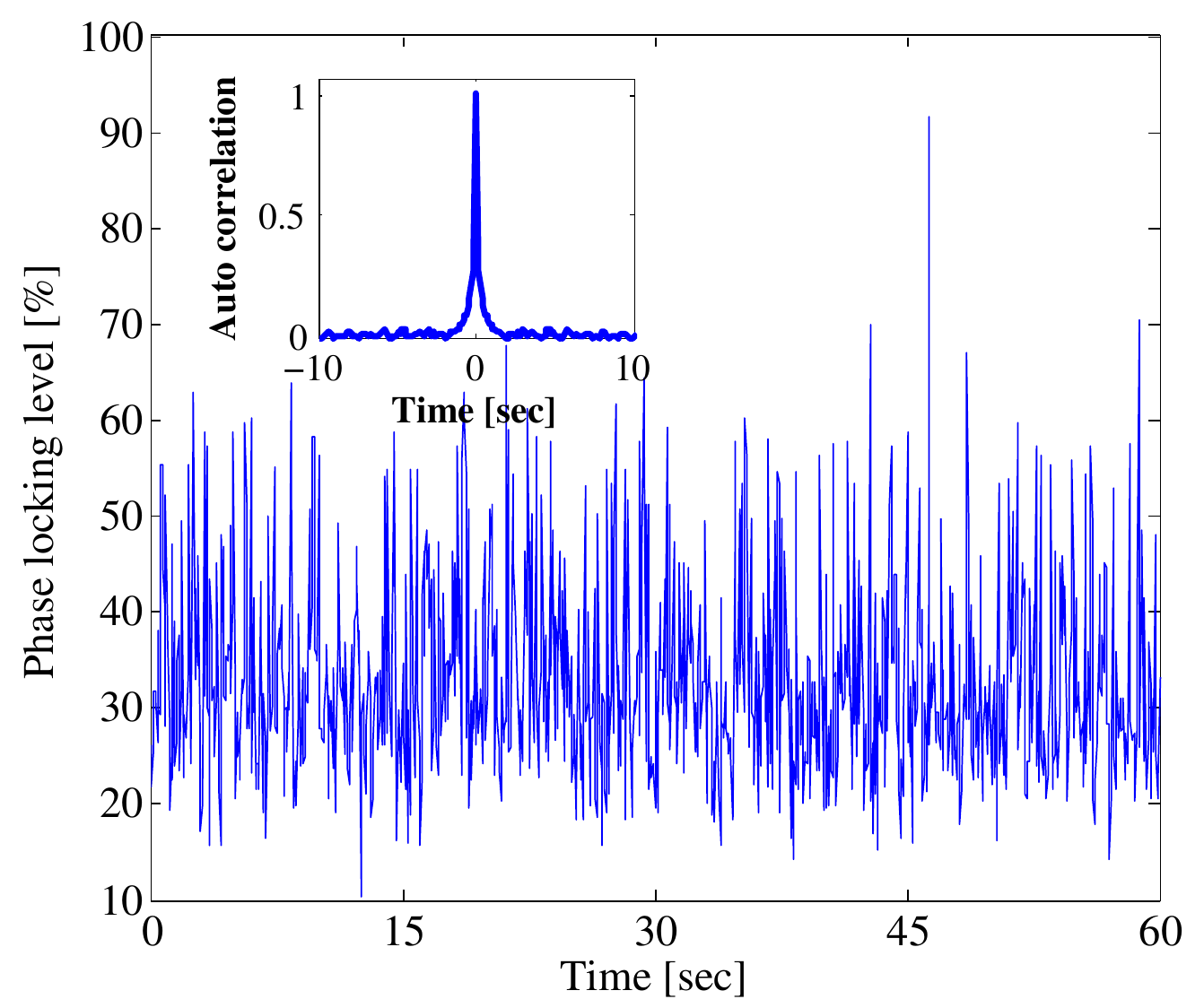}}
\caption{\label{Xcor}Phase locking level as a function of time for
25 fiber lasers. Inset - auto correlation of measurement.}
\end{figure}

\begin{figure}[htb]
\centerline{\includegraphics[width=8cm]{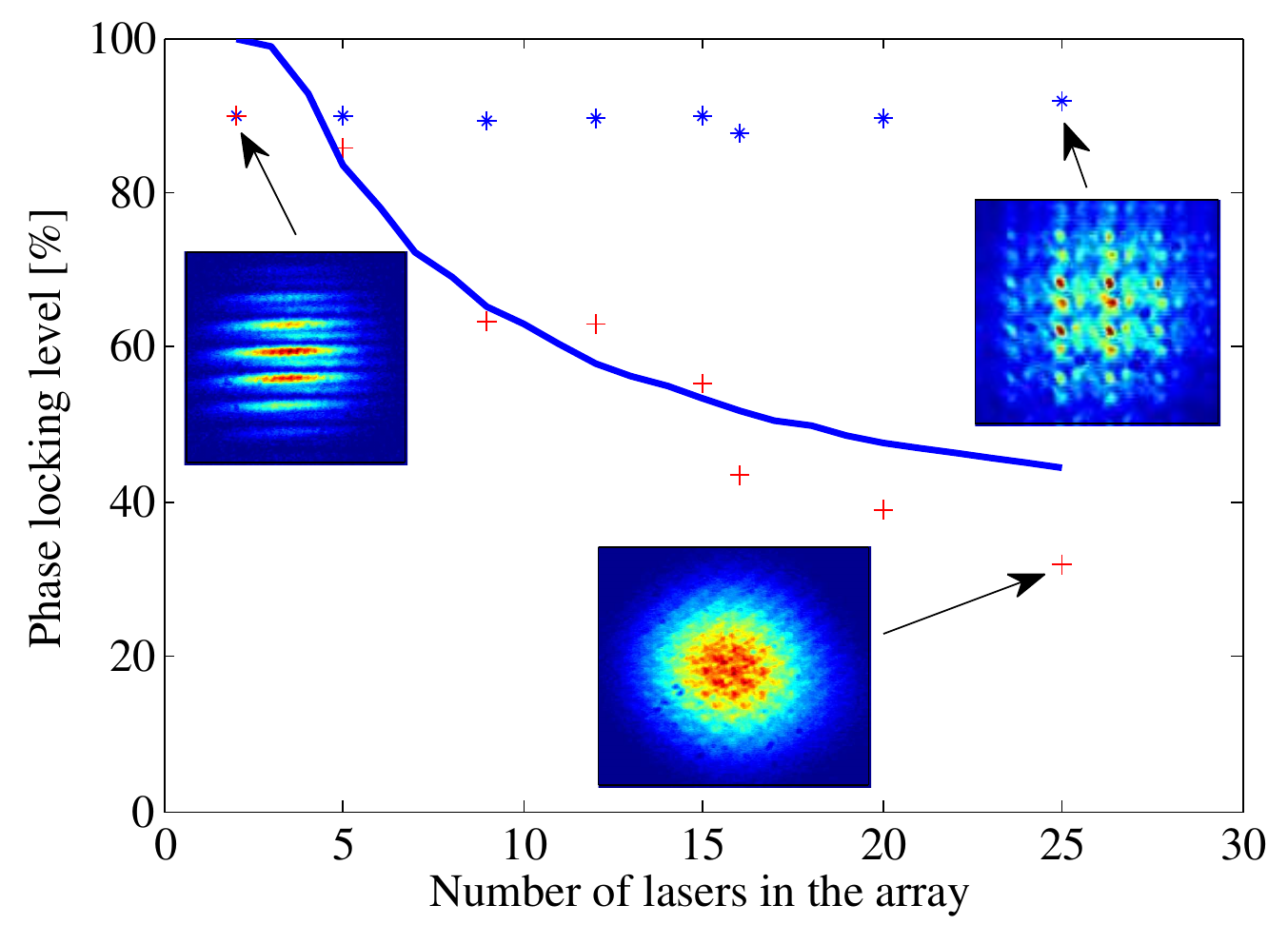}}
\caption{\label{arraysize} Experimental and calculated results of
the average and the maximal phase locking levels as a function of
number of lasers in the array. Stars (blue) - maximal phase locking
level; crosses (red) - average phase locking level; solid curve -
calculated average phase locking level using effective reflectivity
model.}
\end{figure}

\begin{figure}[htb]
\centerline{\includegraphics[width=8cm]{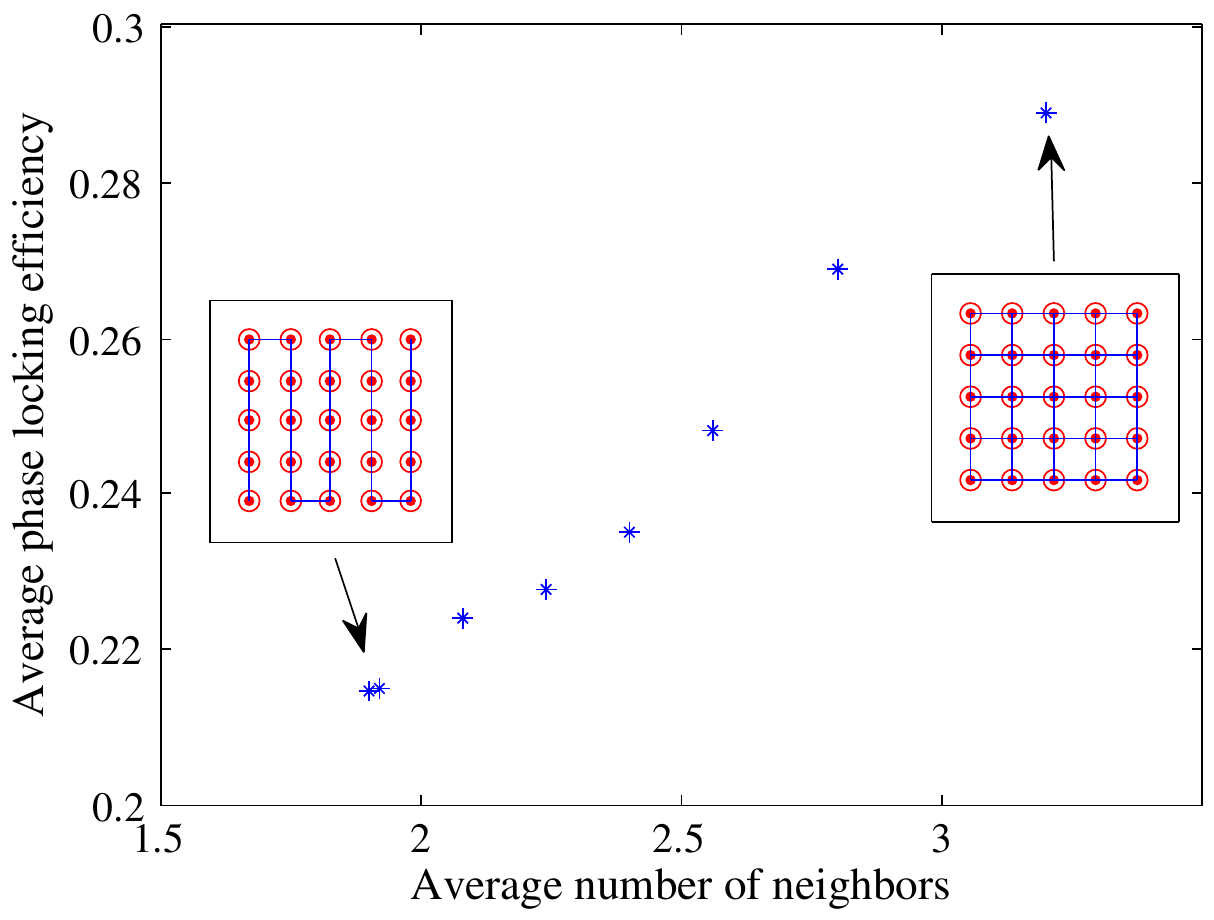}}
\caption{\label{connect}Experimental results of the phase locking
level for 25 fiber lasers as a function of the average number of
coupled neighbors to each laser.}
\end{figure}

Using the configuration shown in Fig.~\ref{setup}, we measured the
phase locking level as a function of time for different number of
lasers in the array and for different connectivities. This was done
by continuously detecting the far-field intensity distribution of
the total output light from the array with a CCD camera and
calculating the average fringe visibility along the x and y
directions, namely the phase locking level that range from 0\% to
100\%, over a period of 10 hours to obtain about $300,000$
measurements. These measurements were then repeated for different
number of lasers in the array and for different connectivities. The
results are presented in Figs.~\ref{Xcor} -~\ref{connect}.
Figure~\ref{Xcor} shows 650 representative measurements of the phase
locking level as a function of time, for 25 lasers with the 2D
connectivity shown in Fig~\ref{connectAll}(i). As seen, the phase
locking level rapidly fluctuates with essentially no correlation
between adjacent measurements, as ascertained in the auto
correlation plot shown in the inset~\cite{FanFluctutuate}. A
striking feature seen in Fig.~\ref{Xcor} is that although the
average phase locking level is rather low (30\%) there are
occasionally high spikes which we denote as the maximal phase
locking level.

Figure~\ref{arraysize} shows the average phase locking level (red
crosses) and the maximal phase locking level (blue stars) as a
function of the number of lasers in the array for the 2D
connectivity. As predicted, the average phase locking level indeed
drops as the number of lasers increases~\cite{no_more_siegman,
no_more_shirakawa, no_more_rothenberg, no_more_shakir}. However, the
instantaneous maximal phase locking level remains high regardless of
the number of lasers in the array. Figure~\ref{arraysize} also
includes representative far-field intensity distributions of two
fiber lasers (inset a), of 25 fiber lasers with instantaneous
maximal phase locking level (inset b), and of 25 fiber lasers with
average low phase locking level (inset c). These results indicate
that as the number of lasers in the array increases, the probability
to find common longitudinal modes rapidly drops as predicted.
However, since the length of each fiber laser fluctuates randomly
(modulus $\lambda$) due to thermal and acoustic variations, there is
a certain probability for instantaneously obtaining a common
longitudinal mode for all fiber lasers. We found that the
probability for obtaining phase locking levels above 90\% drops
rapidly. Specifically, the probability is 0.1\% for 12 lasers,
0.012\% for 16 lasers, 0.004\% for 20 lasers and 0.001\% for 25
lasers.

We calculated the effective reflectivity as a function of number of
lasers in the array~\cite{FridmanOL}. Specifically, we calculated
the maximal effective reflectivity within the bandwidth of our FBG
as a function of the number of coupled fiber lasers whose lengths
were randomly distributed between $2.8m$ to $3.2m$. Then we repeated
the calculations 1000 times, each time choosing a different random
realization of the fiber lasers lengths, and determined the average
of the results. These provide a measure of the phase locking level
for lasers with many longitudinal modes~\cite{motiCommonLM}, and are
plotted as the solid curve in Fig.~\ref{arraysize}. The agreement
between the experimental and calculated results indicates that the
reduction of the average phase locking level is related to the
random cavity lengths of the fiber lasers in the array.

We also determined how the average phase locking level is related to
the connectivity of the fiber lasers in the array. Specifically, we
measured the average phase locking level of an array of 25 fiber
lasers as a function of the average number of coupled neighbors to
each fiber laser, for different coupling connectivities. The results
are presented in Fig.~\ref{connect}. We started with 1D connectivity
of the full array, shown at the left inset, in which the average
number of coupled neighbors to each fiber laser is only $1.9$. Then,
we varied the connectivity and increased the average number of
coupled neighbors and measured the average phase locking level of
the array in each case, up to a 2D connectivity where the average
number of coupled neighbors is $3.2$. As evident, there is a
monotonic increase in the average phase locking level of the array
from $21\%$ up to $29\%$. These results manifest that connectivity
influences the phase locking level, consistent with the expected
increase of an order parameter with dimensionality for all coupled
oscillators~\cite{Strogatz}.

To conclude, we showed that while the average phase locking level
drops as the number of lasers in an array increases, there are still
rare events with high phase locking levels of over 90\%. These
rather rare events have no immediate practical use, but are
significant as a first conclusive proof for the ability to phase
lock a large number of fiber lasers whose cavity lengths are
matched. We also showed that the average phase locking level
increases when increasing the connectivity, suggesting that by
further increasing the connectivity in the array, even higher phase
locking efficiencies may be obtained.

This research was supported in part by the USA-Israel Binational
Science Foundation.

%\pagebreak

\end{document}